\documentclass[11pt,a4paper]{article}

\usepackage[truedimen,margin=30mm]{geometry} 

\usepackage{mathrsfs}
\usepackage{amssymb}
\usepackage{amsmath}
\usepackage{ascmac}
\usepackage{amsthm}
\usepackage[dvipdfmx]{graphicx}
\usepackage{booktabs}
\usepackage{setspace}
\usepackage{natbib}
\usepackage{color}
\usepackage[,
	colorlinks  = true, 
	anchorcolor = black,
	linkcolor   = blue,
	citecolor   = blue, 
	filecolor   = black, 
	urlcolor    = blue
]{hyperref}
\usepackage{bm}
\usepackage{comment}

\usepackage{here}
\usepackage{times}
\usepackage{algorithm}
\usepackage{algorithmic}

\usepackage{enumitem}
\setlist[enumerate, 1]{
  left     = 0em,       % 左余白（インデント）
  itemsep  = -0.5em,     % 項目間のスペース
  topsep   = 0.0em,     % リストと本文の間のスペース（前後）
  labelsep = 0.5em,     % 番号とテキストの間のスペース
  label    = \arabic*.  % 番号のフォーマット（1. 2. 3. ...）
}

\setlist[enumerate, 2]{
  left     = -0.5em,      % 左余白（インデント）
  itemsep  = -0.5em,    % 項目間のスペース
  topsep   = -0.5em,    % リストと本文の間のスペース（前後）
  labelsep = 0.5em,    % 番号とテキストの間のスペース
}

\usepackage{titlesec}
\titleformat*{\section}{\large\bfseries}
\titleformat*{\subsection}{\it}

\def\ep{{\varepsilon}}

\def\bb{\bm{b}}
\def\bc{\bm{c}}
\def\bg{\bm{g}}
\def\bem{\bm{m}}
\def\bu{\bm{u}}
\def\bx{\bm{x}}

\def\bB{\bm{B}}
\def\bC{\bm{C}}
\def\bD{\bm{D}}
\def\bG{\bm{G}}
\def\bH{\bm{H}}
\def\bQ{\bm{Q}}
\def\bX{\bm{X}}
\def\bbe{\bm{\beta}}
\def\bga{\bm{\gamma}}
\def\bka{\bm{\kappa}}

\def\bmu{\bm{\mu}}
\def\bep{\bm{\varepsilon}}
\def\bxi{\bm{\xi}}
\def\bom{\bm{\omega}}
\def\bLa{\bm{\Lambda}}
\def\bPs{\bm{\Psi}}
\def\bSi{\bm{\Sigma}}
\def\bOm{\bm{\Omega}}

\def\Bin{{\rm Bin}}
\def\PG{{\rm PG}}

\def\Unif{{\rm Unif}}

\def\tr{{\rm tr}}

\title{{\bf Robust Spatio-Temporal Distributional Regression}}

\date{}
\author{}

\begin{document}

\maketitle
\doublespacing

\vspace{-1.5cm}
\begin{center}
{\large 
Tomotaka Momozaki$^1$, Shonosuke Sugasawa$^2$, Tomoyuki Nakagawa$^3$, \\Hiroko Kato Solvang$^4$ and Sam Subbey$^{5,6}$ 
}
\end{center}

\noindent
$^1$Department of Information Sciences, Tokyo University of Science \\
$^2$Faculty of Economics, Keio University \\
$^3$School of Data Science, Meisei University\\
$^4$Marine Mammals Research Group, Institute of Marine Research\\
$^5$Research Group on Fisheries Dynamics, Institute of Marine Research\\
$^6$Western Norway University of Applied Sciences

\vspace{5mm}
\begin{center}
{\bf \large Abstract}
\end{center}
Motivated by investigating spatio-temporal patterns of the distribution of continuous variables, we consider describing the conditional distribution function of the response variable incorporating spatio-temporal components given predictors. 
In many applications, continuous variables are observed only as threshold-categorized data due to measurement constraints. 
For instance, ecological measurements often categorize sizes into intervals rather than recording exact values due to practical limitations. 
To recover the conditional distribution function of the underlying continuous variables, we consider a distribution regression employing models for binomial data obtained at each threshold value. 
However, depending on spatio-temporal conditions and predictors, the distribution function may frequently exhibit boundary values (zero or one), which can occur either structurally or randomly. 
This makes standard binomial models inadequate, requiring more flexible modeling approaches. 
To address this issue, we propose a boundary-inflated binomial model incorporating spatio-temporal components. 
The model is a three-component mixture of the binomial model and two Dirac measures at zero and one. 
We develop a computationally efficient Bayesian inference algorithm using P\'olya-Gamma data augmentation and dynamic Gaussian predictive processes. 
Extensive simulation experiments demonstrate that our procedure significantly outperforms distribution regression methods based on standard binomial models across various scenarios.


%\newpage
%%%---------------------------------------------------%%%
%%%         Introduction                              %%%
%%%---------------------------------------------------%%%
\section{Introduction}
Substantial work on statistical modeling of spatio-temporal datasets has been developed, contributing to research and applications across diverse fields \citep{banerjee2014hierarchical, cressie2011statistics, schabenberger2017statistical, stein1999interpolation}. 
Many of these models focus on providing accurate inference for the conditional mean and quantile of a response variable given predictors \citep{finley2012bayesian, gelfand2005spatial, reich2011bayesian, stroud2001dynamic, zuur2009mixed}. 
However, in many applications, researchers are interested in understanding how the entire conditional distribution of the response varies across space and time rather than just its central tendency or specific quantiles. 
This distributional perspective is particularly relevant in various scientific contexts where understanding how entire distributions change across space and time is crucial. 
For instance, in marine ecology, researchers are interested in how the size distribution of fish populations varies spatially and temporally in response to environmental factors such as sea surface temperature and fishing pressure \citep{tu2018fishing}. 
In environmental science, understanding how precipitation distributions shift across regions and seasons is essential for water resource management and climate adaptation planning \citep{konapala2020climate, sharif2025changes, zhang2022temporal}. 
Similarly, in economics, analyzing how income distributions evolve spatially and temporally helps inform regional development policies and inequality reduction strategies \citep{rinz2023re, santos2022regional}. 
Therefore, developing methodologies to estimate conditional distributions at specific locations and time periods is essential for comprehensive spatio-temporal analysis. 

Unlike ordinary regression, which focuses on modeling the conditional mean or quantiles, distributional regression allows us to model how the entire response distribution changes with predictors \citep{klein2024distributional, umlauf2018primer}. 
In this paper, we employ a special class of distributional regression, {\it distribution regression} (DR) proposed by \cite{foresi1995conditional}. 
Our choice of DR is motivated by two key considerations. 
First, in many applications, continuous variables are observed only as threshold-categorized data due to measurement constraints or data collection limitations. 
For instance, ecological measurements often categorize sizes into intervals rather than recording exact values due to practical limitations \citep{weerarathne2021sample}. 
This makes distributional regression inapplicable to continuous data (e.g., density regression). 
Second, DR simply estimates and integrates a sequence of binomial models based on generalized linear models, making the interpretation of distributional changes due to predictors straightforward. 
In addition, estimates in the binomial models are computationally scalable. 

Despite its tractability, DR faces a fundamental limitation of the standard binomial model. 
The standard binomial model assumes a unimodal distribution around the expected value, making it inadequate for modeling discrete proportion data that frequently exhibit boundary values (zero and one) relative to what the standard binomial distribution would predict. 
Such boundary inflation can occur either structurally (true absence/presence of the phenomenon) or randomly (due to sampling variability). 
When spatio-temporal conditions and predictors lead to frequent boundary values in the distribution function, standard binomial models provide poor fits and unreliable inferences. 
We refer to such data as {\it boundary-inflated binomial} (BIB) data. 
The challenge of modeling data with excess boundary values has received considerable attention in the statistical literature. 
The most influential approach for excess zero count data is the zero-inflated Poisson model \citep{lambert1992zero}, which distinguishes structural zeros from random zeros using a two-component mixture: a point mass at zero and a standard Poisson distribution. 
The mixing proportion is modeled using binary regression (e.g., logistic regression). 
This mixture modeling framework has inspired various zero-inflated models, including zero-inflated binomial \citep{hall2000zero} and negative binomial models \citep{ghosh2006bayesian, neelon2019bayesian}. 
Extending beyond zero inflation, \cite{deng2015score} and \cite{tian2015generalized} proposed boundary-inflated binomial (BIB) models as three-component mixtures combining a standard binomial distribution with point masses at both zero and one. 
The mixing proportions in BIB models are typically specified through multinomial regression (e.g., multinomial logistic regression). 
While these boundary-inflated binomial modeling approaches represent important methodological advances, their extension to spatio-temporal data presents additional computational and modeling challenges that require specialized consideration. 
Although extensive research exists on spatio-temporal models, their application to boundary-inflated binomial data remains largely unexplored. 
\cite{gelfand2005spatial} proposed a class of dynamic spatio-temporal models, and \cite{finley2022spnngp} discusses spatial binomial models. 
For continuous proportional data with support in $(0,1)$ (excluding the boundary values 0 and 1), \cite{lagos2017geostatistical} and \cite{ahmadi2025robust} developed spatial beta regression models, while \cite{tang2023zero} introduced spatial extensions to handle structural zeros. 
For non-spatial settings, \cite{li2018efficient} proposed the boundary-inflated beta regression model. 
Recently, \cite{lee2025scalable} proposed continuous binomial (cobin) and mixture of cobin (micobin) regression models for continuous proportional data with support on $[0,1]$. 
While micobin regression can handle boundary values through positive density at zero and one, it addresses non-structural boundary values rather than the structural boundary inflation that we consider. 
Their approach also includes spatial extensions using latent Gaussian processes. 

Previous work on Bayesian approaches to spatio-temporal zero-inflated count data includes \cite{wang2015bayesian} and \cite{sugasawa2023dynamic}, who used zero-inflated Poisson models, and \cite{neelon2019bayesian}, who used the zero-inflated negative binomial model. 
A critical challenge in spatio-temporal modeling arises when survey points are sparse and sampling locations vary over time---a common situation in ecological surveys and environmental monitoring \citep{chen2023relationship, henrys2024adaptive, sevellec2025effect}. 
This irregular sampling design necessitates flexible methodological approaches that can accommodate the varying spatial structure across time periods. 
\cite{wang2015bayesian} address this challenge by dividing the survey area into equally-sized grids and assuming consistent sampling locations throughout the survey period. 
However, their approach requires practitioners to specify grid structures that may not align with the actual spatial sampling design, potentially leading to suboptimal modeling. 
\cite{sugasawa2023dynamic} propose a more flexible approach that explicitly accommodates different spatial locations at each time period by combining dynamic models with Gaussian predictive processes \citep{Banerjee2008}. 
Their method handles irregular sampling designs without requiring pre-specified spatial grids, making it more suitable for practical applications where sampling locations naturally vary over time. 

Our contribution in this paper is to develop a robust spatio-temporal distributional regression (RSTDR) framework that addresses the boundary inflation problem in distribution regression by extending boundary-inflated binomial models to spatio-temporal settings. 
Specifically, we propose a dynamic Gaussian predictive process BIB (DGPP-BIB) model that incorporates dynamic Gaussian predictive processes to handle irregularly spaced sampling locations that vary over time, combining the flexibility of boundary-inflated binomial modeling with computationally efficient inference. 
We develop a scalable Markov Chain Monte Carlo algorithm, specifically a Metropolis-within-Gibbs sampler, that employs Gaussian predictive processes and P\'olya-Gamma data augmentation to achieve tractable posterior computation for the complex hierarchical model structure. 
Through extensive simulation studies, we demonstrate that our RSTDR approach significantly outperforms existing distribution regression methods based on standard binomial models, particularly in scenarios where boundary inflation is prevalent. 
To the best of our knowledge, this is the first work to systematically address boundary-inflated binomial data in spatio-temporal distributional regression settings, providing both theoretical advances and practical computational solutions for this important class of problems. 

The remainder of this paper is organized as follows. 
Section \ref{sec:rstdr} details the RSTDR framework and defines its core, the DGPP-BIB model, along with the computational algorithms for Bayesian inference. 
Section \ref{sec:sim} demonstrates the performance of our approach through extensive simulation studies, comparing it against existing methods across various scenarios. 
Section \ref{sec:concl} concludes with discussions and directions for future research.

%%%--------------------------------------------------%%%
%%%         Methodology                              %%%
%%%--------------------------------------------------%%%
\section{Robust Spatio-Temporal Distributional Regression Framework} \label{sec:rstdr}
This section presents our robust spatio-temporal distributional regression (RSTDR) framework, which extends boundary-inflated binomial models to spatio-temporal settings to address the limitations of standard distribution regression when boundary values are prevalent due to spatio-temporal conditions and predictors. 
We begin with the problem formulation and framework overview (Section \ref{sec:frame}), followed by the detailed DGPP-BIB model specification (Section \ref{sec:dgppbib}), and conclude with the Bayesian inference algorithm (Section \ref{sec:pos}). 

\subsection{Framework Overview and Problem Formulation} \label{sec:frame}
We are interested in the conditional distribution function $F_{it}$ at site $i$ and time $t$ of a response given predictors. 
In our framework, the response is latently continuous, but due to measurement constraints, it is observed as threshold-categorized data. 
That is, suppose that $z_{ijt}^*$ is the latently continuous response variable for $j=1,\ldots,n_{it}$, where $n_{it}$ is the sample size at site $i$ and time $t$, and instead of $z_{ijt}^*$, we observe binomial data consisting of $y_{it}^{(k)} = \sum_{j=1}^{n_{it}} z_{ijt}^{(k)}$ for each threshold value $a_k$, where $z_{ijt}^{(k)} = I(z_{ijt}^*\leq a_k)$, $I(\cdot)$ is the indicator function and $-\infty=a_0<a_1<\ldots<a_K<a_{K+1}=\infty$. 

In DR proposed by \cite{foresi1995conditional}, the values of $F_{it}(z^*)$ at $z^*=a_k$ for $k=1,\ldots,K$, that is, $F_{it}(a_k) = \Pr(z_{ijt}^{(k)}=1)$, are estimated using a sequence of binomial models, $y_{it}^{(k)} \sim \Bin(n_{it}, F_{it}(a_k))$. 
In practice, however, depending on spatio-temporal conditions and predictors, the distribution of $F_{it}(a_k)$ over space and time could frequently exhibit boundary values (zero and one). 
Such boundary inflation can occur either structurally (true absence/presence) or randomly (due to sampling variability), and standard binomial models cannot adequately handle this issue. 
Therefore, the observed proportion data $y_{it}^{(k)}/n_{it}$ have an excess of boundary values, making the standard binomial model inappropriate for such boundary-inflated binomial (BIB) data. 

Modeling individual observations through latent distributions is challenging, as it requires imputing unobserved values and can be sensitive to distributional assumptions. 
Instead, smoothing the distribution function values at each threshold provides a more tractable approach, similar to non-parametric estimation. 
This motivates our distribution regression framework, which we extend to handle boundary inflation in spatio-temporal settings. 

Our RSTDR framework achieves robust spatio-temporal DR for such BIB data by replacing the standard binomial model with our proposed dynamic Gaussian predictive process BIB (DGPP-BIB) model. 
To this end, we develop the DGPP-BIB model in the following section, which incorporates dynamic Gaussian predictive processes to handle boundary inflation and irregular spatio-temporal sampling designs.

%%%-----------------------------------------------------------------------%%%
%%%         DGPP-BIB Model                                                %%%
%%%-----------------------------------------------------------------------%%%
\subsection{Dynamic Gaussian Predictive Process BIB Model} \label{sec:dgppbib} 
Having established the problem formulation, we now develop our dynamic Gaussian predictive process boundary-inflated binomial (DGPP-BIB) model that forms the core of the RSTDR framework. 
This model extends the standard BIB approach to spatio-temporal settings while accommodating irregular sampling locations that vary over time. 

For notational simplicity, let $y_{it}$ and $F_{it}$ denote $y_{it}^{(k)}$ and $F_{it}(a_k)$, respectively. 
Our observed dataset is $(y_{it}, n_{it}, \bm{x}_{it})$, where $y_{it}$ is a binomial response, $n_{it}$ is the sample size of the binomial response, and $\bm{x}_{it}$ is a $q\times1$ predictor vector for $i=1,\ldots,N_t$ and $t=1,\ldots,T$. 
Then, the total number of observation points is $N = \sum_{t=1}^T N_t$. 
We also assume that the location information $\bm{s}_{it}$ is available for each dataset, and in this paper, unless otherwise noted, it is a two-dimensional vector of longitude and latitude. 
Note that this setting allows the sampling locations to be different over $t$. 

To address the boundary inflation problem and distinguish between structural and random boundary occurrences, we model $y_{it}$ as the three component mixtures of the binomial model $\Bin(n_{it}, \pi_{it})$ and two Dirac measures at zero and $n_{it}$. 
Then, $\pi_{it}$ and the mixing probability $p_{kit}$ ($k \in \{0,1\}$) are modeled by the binomial logit and multinomial logit models, respectively, and they include spatio-temporal components $u_{it}$ and $\xi_{kit}$ generated from Gaussian processes.
That is, the overall model is written as 
\begin{equation}\label{eq:stbib}
    \begin{split}
        y_{it} &\sim p_{0it} \delta_{0}(y_{it}) + p_{1it} \delta_{n_{it}}(y_{it}) + (1-p_{0it}-p_{1it}) \Bin(n_{it}, \pi_{it}),  \\
    \pi_{it} &= \frac{e^{\eta_{it}}}{1+e^{\eta_{it}}}, \quad \eta_{it} = \bm{x}_{it}^\top \bm{\beta} + u_{it}, \\
    p_{kit} &= \frac{e^{\psi_{kit}}}{1+e^{\psi_{0it}}+e^{\psi_{1it}}}, \quad \psi_{kit} = \bm{x}_{it}^\top \bm{\gamma}_k + \xi_{kit}, \quad k \in \{0,1\},
    \end{split}
\end{equation}
where $\delta_{a}(y)$ denotes a one-point distribution on $y=a$, and $\bm{\beta}$ and $\bm{\gamma}_k$ are $q\times1$ vectors of coefficients.
Note that the vector of predictors $\bm{x}_{it}$ is not necessarily identical in the binomial logit and multinomial logit models.
The $u_{it}$ and $\xi_{kit}$ are terms for spatio-temporal heterogeneity in $\pi_{it}$ and $p_{kit}$, respectively.
Equivalently, the BIB model \eqref{eq:stbib} can be expressed as follows using the latent indicator variables $\{r_{it}\}$.
\begin{equation} \label{eq:latent}
    y_{it} | r_{it}=k \sim
    \begin{cases}
        \delta_0(y_{it})        & (k=0) \\
        \delta_{n_{it}}(y_{it}) & (k=1) \\
        \Bin(n_{it}, \pi_{it})  & (k=2) 
    \end{cases}, \quad 
    \Pr(r_{it} = k) =
    \begin{cases}
        p_{0it} & (k=0) \\
        p_{1it} & (k=1) \\
        1-p_{0it}-p_{1it} & (k=2) \\
    \end{cases}.
\end{equation}
This representation is useful for deriving the scalable posterior computation, as described in Section \ref{sec:pos}.

One way to feasibly design a model with spatio-temporal heterogeneity is to dynamically model the spatial Gaussian process, as in \cite{gelfand2005spatial} and \cite{finley2012bayesian}. 
However, as noted in the Introduction, when sampling locations vary over time, dynamically modeling spatial Gaussian processes is not straightforward. 
Following \cite{sugasawa2023dynamic}, we address this problem using Gaussian predictive processes \citep{Banerjee2008}, a special class of low rank approximation that projects a spatial Gaussian process to a lower dimensional subspace, at each time point to represent the time variability of the approximated spatial Gaussian process, referred to as the dynamic Gaussian predictive process (DGPP). 

Formally, let $\{ \bm{s}_1,\ldots,\bm{s}_{N_t} \}_{t=1,\ldots,T}$ and $\{ \bar{\bm{s}}_1,\ldots,\bar{\bm{s}}_M \}$ be sets of sampled locations and knots over the region, respectively, the DGPP is expressed as
\begin{equation}\label{eq:dgpp}
    \begin{aligned}
        \bm{u}_t &= \bar{\bm{c}}_t (\bm{\theta}_u)^\top \bar{\bm{C}}(\bm{\theta}_u)^{-1} \bar{\bm{u}}_t, \quad & 
        \bar{\bm{u}}_t &| \bar{\bm{u}}_{t-1} \sim N_M(\bar{\bm{u}}_{t-1}, \tau_u^{-1} \bar{\bm{C}}(\bm{\theta}_u)), \\
        \bm{\xi}_{kt} &= \bar{\bm{c}}(\bm{\theta}_{\xi_k})^\top \bar{\bm{C}}(\bm{\theta}_{\xi_k})^{-1} \bar{\bm{\xi}}_{kt}, \quad &
        \bar{\bm{\xi}}_{kt} &| \bar{\bm{\xi}}_{k,t-1} \sim N_{M}(\bar{\bm{\xi}}_{k,t-1}, \tau_{\xi_k}^{-1} \bar{\bm{C}}(\bm{\theta}_{\xi_k}))
    \end{aligned}
\end{equation}
for $t=1,\ldots,T$, where $\bm{u}_t = (u_{1t},\ldots,u_{N_t t})^\top$ and $\bm{\xi}_{kt} = (\xi_{k1t},\ldots,\xi_{kN_tt})^{\top}$ are the $N_t \times 1$ approximated spatial component vectors, $\bar{\bm{u}}_t = (\bar{u}_{1t},\ldots,\bar{u}_{Mt})^\top$ and $\bar{\bm{\xi}}_{kt} = (\bar{\xi}_{k1t},\ldots,\bar{\xi}_{kMt})^\top$ are the $M\times 1$ lower dimensional vectors ($M < N_t$), $\bar{\bm{c}}_t (\bm{\theta})$ is the $M\times N_t$ matrix whose $(m,i)$-element is a valid correlation function $\rho(\bar{\bm{s}}_m, \bm{s}_i; \bm{\theta})$, $\bar{\bm{C}}(\bm{\theta})$ is the $M\times M$ matrix with $(m,m')$-element being $\rho(\bar{\bm{s}}_m, \bar{\bm{s}}_{m'}; \bm{\theta})$, and $\tau_u$ and $\tau_{\xi_k}$ are the spatial precision components. 
We customarily specify $\rho(\bm{s}_i, \bm{s}_{i'};\bm{\theta}) = \rho(\|\bm{s}_i-\bm{s}_{i'}\|;\phi)$ with spatial range parameter $\phi$ such as exponential correlation function $\exp(-\|\bm{s}_i-\bm{s}_{i'}\|/\phi)$. 
As prior distributions, we assume $\bar{\bm{u}}_0 \sim N_M(\bm{0}, \bar{\bm{C}}(\bm{\theta}_u))$ and $\bar{\bm{\xi}}_{k0} \sim N_M(\bm{0}, \bar{\bm{C}}(\bm{\theta}_{\xi_k}))$. 
Note that we determine the locations of knots $\{ \bar{\bm{s}}_1,\ldots,\bar{\bm{s}}_M \}$ by applying the $k$-means clustering to sampled locations $\{ \bm{s}_1,\ldots,\bm{s}_{N_t} \}_{t=1,\ldots,T}$ \citep{ver2015estimating, sugasawa2023dynamic}. 

Therefore, the proposed model consists of the mixture distribution for $y_{it}$, given in (\ref{eq:stbib}), with latent dynamic spatial processes, defined in \eqref{eq:dgpp}, to which we refer as the dynamic Gaussian predictive process BIB (DGPP-BIB) model.
We consider Bayesian inference on the unknown parameters $\Theta = \{ \bm{\beta}, \{\bm{\gamma}_{k}\}, \tau_u, \{\tau_{\xi_k}\}, \phi_u, \{\phi_{\xi_k}\} \}$ as well as the spatial components $\{\bm{u}_t\}$ and $\{\bm{\xi}_{kt}\}$, and the latent variables $\{r_{it}\}$ by generating random samples from a joint posterior distribution in the DGPP-BIB model. 
From equations \eqref{eq:latent} and \eqref{eq:dgpp}, the full hierarchical model can be written as 
\begin{equation} \label{eq:post}
    \begin{split}
        & p(\Theta) \prod_{t=1}^T\prod_{i=1}^{N_t}
\frac{\{e^{\psi_{0it}}\delta_{0}(y_{it})\}^{I(r_{it}=0)} \{e^{\psi_{1it}}\delta_{n_{it}}(y_{it})\}^{I(r_{it}=1)} \Bin(y_{it};n_{it},\pi_{it})^{I(r_{it}=2)}}{1+e^{\psi_{0it}}+e^{\psi_{1it}}} \\
        & \times \phi_M(\bar{\bu}_0; \bm{0}, \tau_{u}^{-1} \bar{\bC}(\phi_u)) \phi_M(\bar{\bm{\xi}}_{00}; \bm{0}, \tau_{\xi_0}^{-1} \bar{\bC}(\phi_{\xi_0})) \phi_M(\bar{\bm{\xi}}_{10}; \bm{0}, \tau_{\xi_1}^{-1} \bar{\bC}(\phi_{\xi_1})) \\
        & \times \prod_{t=1}^T \phi_M(\bar{\bm{u}}_t; \bar{\bm{u}}_{t-1}, \tau_{u}^{-1} \bar{\bC}(\phi_u)) \phi_M(\bar{\bm{\xi}}_{0t}; \bar{\bm{\xi}}_{0t-1}, \tau_{\xi_0}^{-1} \bar{\bC}(\phi_{\xi_0})) \phi_M(\bar{\bm{\xi}}_{1t}; \bar{\bm{\xi}}_{1t-1}, \tau_{\xi_1}^{-1} \bar{\bC}(\phi_{\xi_1})),
    \end{split}
\end{equation}
where $p(\Theta)$ is the joint prior of $\Theta$, $\Bin(y;n,\pi)$ is the density function of the binomial distribution, and $\phi_d(\bm{y}; \bm{\mu}, \bm{\Sigma})$ is the density function of the $d$-dimensional multivariate normal distribution with mean $\bm{\mu}$ and covariance matrix $\bm{\Sigma}$.

\subsection{Bayesian Inference Algorithm}\label{sec:pos} 
To obtain the marginal posterior distribution for each unknown parameter in $\Theta$, we develop a Markov Chain Monte Carlo algorithm, specifically a Metropolis-within-Gibbs sampler, employing the joint posterior distribution \eqref{eq:post}.
Although the binomial logit likelihood and the multinomial logit seem to yield intractable conditional posterior distributions for the coefficient parameters $\bbe$ and $\bga_{k}$, as well as the Gaussian spatial components $\bu_t$ and $\bxi_k$, the use of P\'olya-gamma data augmentation \citep{polson2013bayesian} facilitates the posterior computation.
Our Metropolis-within-Gibbs sampler algorithm is as follows: 

\medskip 
Let $\bX$, $\bka$, and $\bka_k$ be an $N \times q$ matrix and $N \times 1$ vectors constructed by vertically stacking $\{\bX_t\}$, $\{\bka_t = (\kappa_{1t},\ldots,\kappa_{N_tt})^\top\}$, and $\{\bka_{kt}=(\kappa_{k1t},\ldots,\kappa_{kN_tt})^\top\}$, respectively, and $\kappa_{it} = y_{it} - n_{it}/2$ and $\kappa_{kit} = I(r_{it}=k) - 1/2$. 
$\bOm$ and $\bOm_k$ are $N\times N$ diagonal matrices of $\{\bm{\omega}_t=(\omega_{1t},\ldots,\omega_{N_tt})^\top\}$ and $\{\bom_{kt}=(\omega_{k1t},\ldots,\omega_{kN_tt})^\top\}$, respectively. 
The $MT \times 1$ vectors $\bar{\bu}$ and $\bar{\bxi}_k$ are constructed by vertically stacking $\{\bar{\bu}_t\}$ and $\{\bar{\bxi}_{kt}\}$, respectively, while the $MT \times 1$ vectors $\bar{\bu}_{0MT}$ and $\bar{\bxi}_{k0MT}$ are constructed by vertically stacking $T$ copies of $\bar{\bu}_0$ and $\bar{\bxi}_{k0}$, respectively. 
$\bar{\bD}$ and $\bar{\bD}_k$ denote $N\times MT$ block diagonal matrices of $\{\bar{\bD}_t=\bar{\bc}_t(\phi_u)^\top\bar{\bC}(\phi_u)^{-1}\}$ and $\{\bar{\bD}_{kt}=\bar{\bc}_t(\phi_{\xi_k})^\top\bar{\bC}(\phi_{\xi_k})^{-1}\}$, respectively, and $\bu = \bar{\bD}\bar{\bu}$ and $\bxi_k = \bar{\bD}_k \bar{\bxi}_k$. 
$\bH$ denotes a $T\times T$ matrix with 1s on the main diagonal and $-1$s on the subdiagonal, and we define $\bar{\bC}_H=(\bH^\top\bH)^{-1}\otimes\bar{\bC}(\phi_u)$ and $\bar{\bC}_{H_k}=(\bH^\top\bH)^{-1}\otimes\bar{\bC}(\phi_{\xi_k})$, where $\otimes$ is the Kronecker product. 
The subscript $*$ on any matrix or vector indicates the submatrix or subvector corresponding to entries where $r_{it} = 2$. 

Regarding the prior distributions, we assign $\bbe\sim N(\bb_{0}, \bB_{0}^{-1})$, $\bga_{k} \sim N(\bg_{k0}, \bG_{k0}^{-1})$, $\tau_u\sim {\rm Ga}(a_{u_0}, b_{u_0})$, $\tau_{\xi_k}\sim {\rm Ga}(a_{\xi_{k0}}, b_{\xi_{k0}})$, and $\phi_u, \phi_{\xi_k}\sim {\rm Unif}(\underline{\phi}, \overline{\phi})$, where $\bb_{0}$, $\bB_{0}$, $\bg_{k0}$, $\bG_{k0}$, $a_{u_0}$, $b_{u_0}$, $a_{\xi_{k0}}$, $b_{\xi_{k0}}$, $\underline{\phi}$, and $\overline{\phi}$ are hyperparameters.
Then, in each iteration of the Metropolis-within-Gibbs sampler, 
\begin{enumerate}
    \item 
    Sample $\{r_{it}\}$ from the categorical distribution with the probabilities of each category
    \begin{align*}
        &\tilde{p}_{0it} = \dfrac{e^{\psi_{0it}}\delta_0(y_{it})}{e^{\psi_{0it}}\delta_0(y_{it}) + e^{\psi_{1it}}\delta_{n_{it}}(y_{it}) + \Bin(y_{it};n_{it},\pi_{it})} & (k=0), \\
        &\tilde{p}_{1it} = \dfrac{e^{\psi_{1it}}\delta_{n_{it}}(y_{it})}{e^{\psi_{0it}}\delta_0(y_{it}) + e^{\psi_{1it}}\delta_{n_{it}}(y_{it}) + \Bin(y_{it};n_{it},\pi_{it})} & (k=1), \\
        &1-\tilde{p}_{0it}-\tilde{p}_{1it} & (k=2).
    \end{align*}

    \item 
    \begin{enumerate}
        \item \label{sample_omega} 
        Sample $\{\omega_{it}\}$ from $\PG(n_{it}, \eta_{it})$, where $\eta_{it} = \bx_{it}^\top \bbe + u_{it}$. 

        \item 
        Sample $\bbe$ from $N_q(\bB^{-1}\bb, \bB^{-1})$, where 
        \begin{equation*}
            \bb=\bB_0\bb_0+\bX_*^\top(\bka_*-\bOm_* \bu_*), \quad 
            \bB=\bB_0+\bX_*^\top\bOm_*\bX_*.
        \end{equation*}

        \item \label{sample_u}
        Sample $\bar{\bu}$ from $N_{MT}(\bQ^{-1}\bem, \bQ^{-1})$, where 
        \begin{equation*}
            \bem=\tau_u\bar{\bC}_H^{-1}\bar{\bu}_{0MT}+\bar{\bD}_*^\top(\bka_*-\bOm_*\bX_*\bbe), \quad 
            \bQ=\tau_u\bar{\bC}_H^{-1}+\bar{\bD}_*^\top\bOm_*\bar{\bD}_*. 
        \end{equation*}

        \item 
        Sample $\bar{\bu}_0$ from $N_M(2^{-1}\bar{\bu}_1, (2\tau_{u})^{-1} \bar{\bC}(\phi_u))$. 

        \item 
        Sample $\tau_u$ from ${\rm Ga}(a_u, b_u)$, where 
        \begin{align*}
            a_u &= a_{u_0} + \frac{M(T+1)}{2}, \\ 
            b_u &= b_{u_0} + \frac{1}{2}\left[ \bar{\bu}_0^\top \bar{\bC}(\phi_u)^{-1} \bar{\bu}_0 + (\bar{\bu}-\bar{\bu}_{0MT})^\top \bar{\bC}_H^{-1} (\bar{\bu}-\bar{\bu}_{0MT}) \right]. 
        \end{align*}

        \item 
        Sample $\phi_u$ from 
        \begin{equation*}
            \left[\prod_{t=1}^T \prod_{i=1}^{N_t} \Bin(y_{it};n_{it},\pi_{it})^{I(r_{it}=2)}\right]  \phi_M(\bar{\bu}_0; \bm{0}, \tau_{u}^{-1} \bar{\bC}(\phi_u)) \phi_{MT}(\bar{\bu};\bar{\bu}_{0MT},\tau_u^{-1}\bar{\bC}_H), 
        \end{equation*}
        where $\phi_u \in (\underline{\phi}, \overline{\phi})$, with the random-walk Metropolis-Hastings algorithm. 
    \end{enumerate}

    \item For each $k=0,1$, 
    \begin{enumerate}
        \item \label{sample_omegak}
        Sample $\{\omega_{kit}\}$ from $\PG(1,\psi_{kit}-\Psi_{kit})$, where 
        \begin{equation*}
            \psi_{kit} = \bm{x}_{it}^\top \bm{\gamma}_k + \xi_{kit}, \quad 
            \Psi_{kit} = \log\left\{ 1+\sum_{\ell\in \{0,1\}\setminus \{k\}}\exp(\psi_{\ell it}) \right\}. 
        \end{equation*}

        \item 
        Sample $\bga_k$ from $N_q(\bG_k^{-1}\bg_k, \bG_k^{-1})$, where 
        \begin{equation*}
            \bg_k = \bG_{k0}\bg_{k0} + \bX^\top \{\bka_k + \bOm_k(\bPs_k - \bxi_k)\}, \quad 
            \bG_k = \bG_{k0} + \bX^\top \bOm_k \bX. 
        \end{equation*}

        \item \label{sample_xi}
        Sample $\bar{\bxi}_k$ from $N_{MT}(\bQ_k^{-1}\bem_k,\bQ_k^{-1})$, where 
        \begin{equation*}
            \bem_k = \tau_{\xi_k} \bar{\bC}_{H_k}^{-1} \bar{\bxi}_{k0MT} + \bar{\bD}_k^\top \{\bka_k + \bOm_k (\bPs_k - \bX \bga_k) \}, \quad 
            \bQ_k = \tau_{\xi_k} \bar{\bC}_{H_k}^{-1} + \bar{\bD}_k^\top \bOm_k \bar{\bD}_k, 
        \end{equation*}

        \item 
        Sample $\bar{\bxi}_{k0}$ from $N_{M}(2^{-1}\bar{\bxi}_{k1}, (2\tau_{\xi_k})^{-1}\bar{\bC}(\phi_{\xi_k}))$. 

        \item 
        Sample $\tau_{\xi_k}$ from ${\rm Ga}(a_{\xi_k}, b_{\xi_k})$, where 
        \begin{align*}
            a_{\xi_k} &= a_{\xi_{k0}} + \frac{M(T+1)}{2}, \\
            b_{\xi_k} &= b_{\xi_{k0}} + \frac{1}{2}\left[ \bar{\bxi}_{k0}^\top \bar{\bC}(\phi_{\xi_k})^{-1} \bar{\bxi}_{k0} + (\bar{\bxi}_k-\bar{\bxi}_{k0MT})^\top \bar{\bC}_{H_k}^{-1} (\bar{\bxi}_k-\bar{\bxi}_{k0MT}) \right]. 
        \end{align*}

        \item 
        Sample $\phi_{\xi_k}$ from 
        \begin{equation*}
            \left[\prod_{t=1}^T\prod_{i=1}^{N_t}
            p_{kit}^{I(r_{it}=k)}\right] \phi_M(\bar{\bm{\xi}}_{k0}; \bm{0}, \tau_{\xi_k}^{-1} \bar{\bC}(\phi_{\xi_k})) \phi_{MT}(\bar{\bxi}_k; \bar{\bxi}_{k0MT}, \tau_{\xi_k}^{-1} \bar{\bC}_{H_k}), 
        \end{equation*}
        where $\phi_{\xi_k} \in (\underline{\phi}, \overline{\phi})$, with the random-walk Metropolis-Hastings algorithm. 
    \end{enumerate}
\end{enumerate}

\medskip
\medskip
Steps \ref{sample_u} and \ref{sample_xi} involve sampling spatial components from medium to large-dimensional normal distributions. 
One of the key computational advantages of our DGPP-BIB model is that the dynamic Gaussian predictive process formulation can be transformed into multivariate normal distributions with block tridiagonal precision matrix structures (see Appendix \ref{app:dgpp} for the detailed transformation). 
This transformation enables us to derive Gaussian full conditional posterior distributions that retain the block tridiagonal precision matrix structure for both $\bar{\bu}$ and $\bar{\bxi}_k$ (see the precision matrices $\bQ$ and $\bQ_k$ in Appendix \ref{app:derivative}). 
The block tridiagonal precision matrix structure enables us to implement efficient simulation smoothing algorithms that are computationally superior to traditional Kalman filter-based approaches \citep{carter1994gibbs, fruhwirth1994data, de1995simulation, durbin2002simple}. 
\cite{rue2001fast} developed the Cholesky factor algorithm for efficiently sampling from general Gaussian Markov random fields with band diagonal precision matrices, while \cite{mccausland2011simulation} introduced a simulation algorithm designed for normal linear state-space models that directly exploits block tridiagonal structures. 
These approaches employ joint sampling to avoid the poor MCMC mixing and potential convergence issues associated with one at a time sampling \citep{carlin1992monte} and achieve improved MCMC performance \citep{liu1994covariance, liu1994collapsed}, while eliminating or significantly reducing the computational costs of recursive conditional distributions $p(\bar{\bu}_t | \bar{\bu}_{t+1}, -)$ required by Kalman filter-based methods. 
These precision-based methods provide flexible options for efficiently sampling $\bar{\bu}$ and $\bar{\bxi}_k$ depending on factors such as the dimensions $M$ and $T$, and the implementation language.
The detailed sampling procedures for both approaches are provided in Appendix \ref{app:spatial}. 

It should be noted that all the above sampling steps are simply generating from some familiar distributions, except spatial range parameters $\phi_u$ and $\{\phi_{\xi_k}\}$, so that no rejection steps are require in generating posterior samples from the full conditional distributions, which would prevent high serial correlations of the posterior samples. 
In Steps \ref{sample_omega} and \ref{sample_omegak}, we adopt R package, \verb+pgdraw+, to sample from the P\'olya-gamma distribution. 
The detailed derivation of all full conditional posterior distributions is provided in Appendix \ref{app:derivative}.

%----------------------------------------%
%              Simulation                %
%----------------------------------------%
\section{Simulation study}\label{sec:sim}
This section conducts simulation studies to evaluate and compare the performance of our proposed RSTDR framework against existing distribution regression methods for recovering spatio-temporal distribution functions evaluated at some thresholds under various scenarios with boundary inflation.  
In this study, we set $T=10$, $N_t=50$, and generated $n_{it}$ from the uniform distribution on $(50, 100)$. 
The spatial location, $s_{it}=(s_{it1}, s_{it2})$ are generated from the uniform distribution on $(-1,1)\times (-1,1)$, and univariate predictor $x_{it}$ is generated from $N(0, 0.5^2)$. 
For the distribution of latently continuous response variable, we consider three component mixtures of the log-normal distribution and two uniform distributions, that is, 
\begin{equation*}
    F_{it}(z^*|x) = 
    \begin{cases}
        \Unif(z^*; a_K, a_K+c) & \mbox{with probability } \lambda_{0it} \\ 
        \Unif(z^*; 0, a_1) & \mbox{with probability } \lambda_{1it} \\ 
        {\rm LN}(z^*;\mu_{it}(x), \sigma_{it}^2(x)) & \mbox{with probability } 1-\lambda_{0it}-\lambda_{1it}
    \end{cases}
\end{equation*}
where ${\rm LN}(z; \mu, \sigma^2)$ denotes the log-normal distribution function with log-mean $\mu$ and log-variance $\sigma^2$, $\Unif(z; a, b)$ denotes the uniform distribution function with lower bound $a$ and upper bound $b$, and $c$ is a constant value. 
For $k \in \{0,1\}$, $\lambda_{kit} = \exp(\nu_{kit})/\sum_{k'=0}^2 \exp(\nu_{k'it})$, where 
\begin{align*}
    \nu_{0it} = -1   +0.5x_{it} + \zeta^{(0)}(s_{it}) + \iota_t^{(0)}, \quad 
    \nu_{1it} = -1.5 -   x_{it} + \zeta^{(1)}(s_{it}) + \iota_t^{(1)}, 
\end{align*}
and $\nu_{2it}=0$, and for $\mu_{it}\equiv \mu_{it}(x)$ and $\sigma_{it}\equiv \sigma_{it}(x)$, we adopt the following form: 
\begin{equation*}
    \mu_{it}    = 1+x_{it} + \zeta^{(2)}(s_{it}) + \iota_t^{(2)}, \quad 
    \sigma_{it} = \exp(-1.5+0.2x_{it} +0.5 \zeta^{(2)}(s_{it}) + 0.5\iota_t^{(2)}), 
\end{equation*}
where $\zeta^{(k)}(s_{it})$ is a spatial effect and $\iota_t^{(k)}$ is a time effect. 
The time effect is defined as 
\begin{align*}
    \iota_t^{(0)}=\frac12\sin\left(\frac{\pi t}{2}\right), \ \ \ \ 
    \iota_t^{(1)}=-\frac12\cos\left(\frac{\pi t}{2}\right), \ \ \ \ 
    \iota_t^{(2)}=\frac{1.5t}{T}, 
\end{align*}
for $t=1,\ldots,T$. 
Regarding the spatial effect, we adopt the following two scenarios: 
\begin{align*}
    {\rm (Scenario\ 1)} \quad \quad
    &\zeta^{(0)}(s_{it})=\sin(s_{it1}), \\ 
    &\zeta^{(1)}(s_{it})=\cos(s_{it1}), \\
    &\zeta^{(2)}(s_{it})=\exp(-2s_{it1}^2-2s_{it2}^2) + s_{it1} + s_{it2}. \\
    {\rm (Scenario\ 2)} \quad \quad
    &\zeta^{(0)}(s_{it})=\sin(s_{it1})-\frac12 I(s_{it2}>0), \\ 
    &\zeta^{(1)}(s_{it})=\cos(s_{it1})-\frac12 I(s_{it2}>0), \\
    &\zeta^{(2)}(s_{it})=\exp(-2s_{it1}^2-2s_{it2}^2) + 2I(s_{it1} + s_{it2}>0)-1.
\end{align*}
For each $i$ and $t$, we generate $n_{it}$ random samples from the conditional distribution $F_{it}(z^*|x)$, and calculate the number of samples included in the interval $(a_{k-1}, a_{k}]$ for $k=1,\ldots, K$. 
Here, the threshold $a_k$ is set as $K=7$, $(a_1, \ldots,a_K)=(1,2,4,6,8,10,14)$, $a_0=0$ and $a_{K+1}=\infty$. 

For the simulated dataset, we apply the proposed RSTDR method to recover the value of distribution function at each threshold. 
For comparison, we also apply the binomial (BN) spatio-temporal model that uses the same DGPP framework as our proposed method but without boundary inflation components. 
Furthermore, as more simple approaches, we adopted two additional methods. 
For generalized additive models (GAM), we used the \verb+mgcv+ package with binomial family and logistic link function, applying smooth functions to each predictor. 
For extreme gradient boosting tree (XGB), we used the \verb+xgboost+ package with default settings for binary classification. 
Both GAM and XGB methods use the four-dimensional variables $(x_{it}, s_{it1}, s_{it2}, t)$ as input and the binomial observation as output. 
For BIB and BN, we generated 2000 posterior samples after discarding the first 1000 samples as burn-in, to compute posterior means and 95\% credible intervals of $F_{it}(a_k)$. 
For XGB, the number of trees is set to 1000. 

\begin{figure}[htb!]
\centering
\includegraphics[width=\columnwidth]{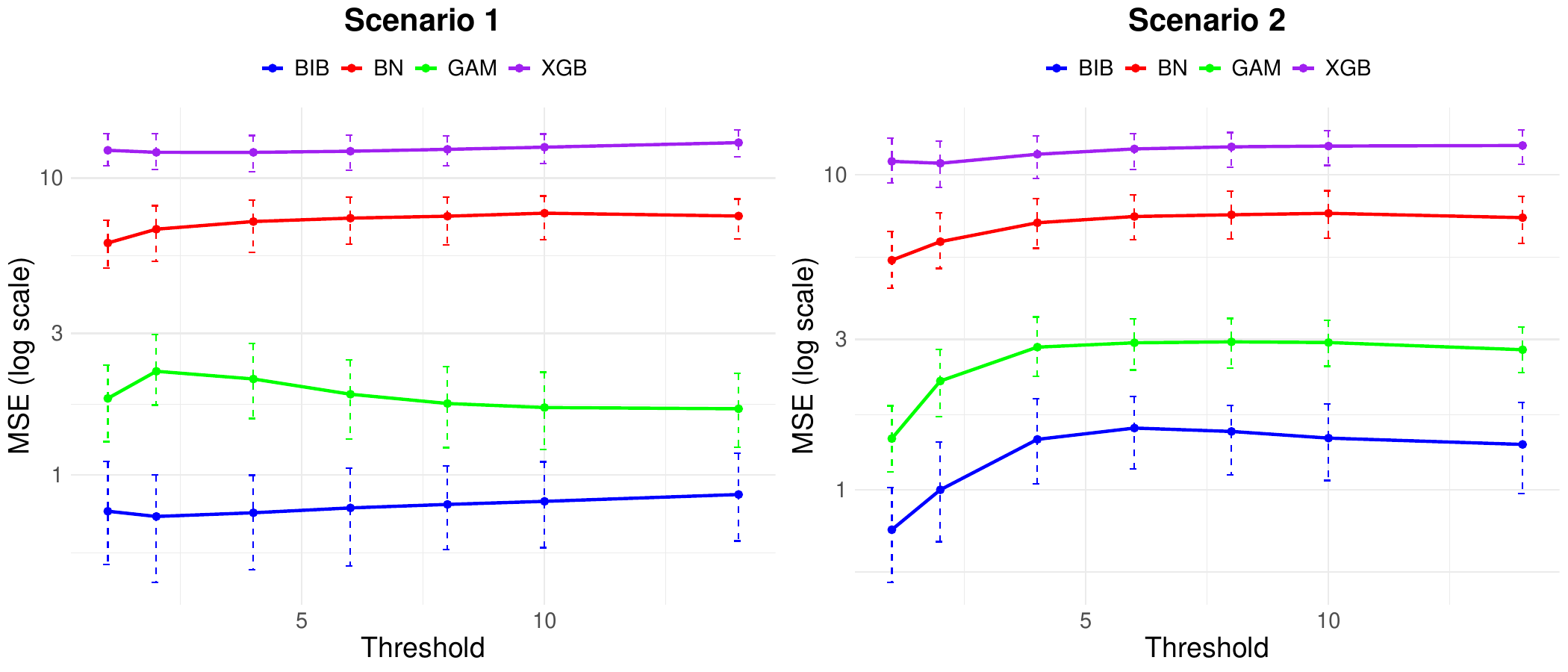}
\caption{
Mean squared errors (MSE) of the point estimators of $F(a_k)$ for seven thresholds, $a_k\in \{1,2,4,6,8,10,14\}$, obtained from the four methods, based on 100 replicated datasets. 
The vertical lines correspond to $95\%$ interval of MSE among 100 replications.
}
\label{fig:sim}
\end{figure}

For $r=1,\ldots,R(=100)$ replicated dataset, we evaluated the mean squared errors (MSE) at each $a_k$, defined as 
$$
{\rm MSE}_k^{(r)}=\frac{1}{\sum_{t=1}^T N_t}\sum_{t=1}^{T}\sum_{i=1}^{N_t} \left\{ \widehat{F}_{it}^{(r)}(a_k)-F_{it}^{(r)}(a_k) \right\}^2,\ \ \ \ k=1,\ldots,K,
$$
where $\widehat{F}_{it}^{(r)}(a_k)$ and $F_{it}^{(r)}(a_k)$ are estimated and true values of $F_{it}(a_k)$, respectively, in the $r$th replication. 
For each $k$, the replicated MSE is summarized by its mean and $95\%$ intervals by computing the lower $2.5\%$ quantile and upper $2.5\%$ quantile of $\{{\rm MSE}_k^{(1)},\ldots, {\rm MSE}_k^{(R)}\}$. 
In the following simulation results, we refer to our proposed RSTDR framework as ``BIB'' (boundary-inflated binomial) and the comparison method without boundary inflation as ``BN'' (binomial) for clarity in figures and tables. 
The results are presented in Figure \ref{fig:sim}. 
The proposed BIB method consistently achieves the lowest MSE across all threshold values in both scenarios, demonstrating superior performance compared to the three alternative approaches. 
Notably, the BN method shows substantially higher MSE than BIB, highlighting the critical importance of incorporating boundary inflation components in the model. 

% Figure 

Regarding the interval estimation, we evaluated the coverage probability (CP) and average length (AL), defined as 
\begin{align*}
&{\rm CP}_k=\frac{1}{R\sum_{t=1}^T N_t}\sum_{r=1}^{R}\sum_{t=1}^{T}\sum_{i=1}^{N_t} I\Big(F_{it}^{(r)}(a_k) \in {\rm CI}_{it}^{(r)}(a_k) \Big),\\
&{\rm AL}_k=\frac{1}{R\sum_{t=1}^T N_t}\sum_{r=1}^{R}\sum_{t=1}^{T}\sum_{i=1}^{N_t} |{\rm CI}_{it}^{(r)}(a_k)|,
\end{align*}
for $k=1,\ldots,K$. 
The results are given in Table \ref{tab:sim}. 
The proposed BIB method demonstrates well-calibrated uncertainty quantification with coverage probabilities close to the nominal 95\% level in both scenarios. 
In Scenario 1, coverage probabilities range from 91.1\% to 94.0\% across different thresholds, while in Scenario 2, they range from 89.1\% to 93.3\%. 
The slightly lower coverage in Scenario 2 reflects the increased complexity introduced by spatial discontinuities, yet the values remain reasonably close to the target level. 
In contrast, the BN method shows severely inadequate coverage probabilities in both scenarios, ranging from 7.0\% to 16.6\% in Scenario 1 and 7.9\% to 16.6\% in Scenario 2, representing a dramatic underestimation of uncertainty. 
This substantial deterioration in coverage occurs because the BN model fails to account for the additional variability introduced by boundary inflation, resulting in overly narrow credible intervals with average lengths approximately 3-7 times smaller than those of BIB (e.g., 0.05 vs 0.28 for threshold 1 in Scenario 1). 
The consistently poor coverage across all thresholds and scenarios indicates that ignoring boundary inflation leads to systematic underestimation of parameter uncertainty, making the BN approach unreliable for statistical inference in boundary-inflated settings. 
This finding underscores the critical importance of the boundary inflation components in our RSTDR framework for producing trustworthy uncertainty quantification in spatio-temporal distributional regression. 

%  Table 
\begin{table}[htb!]
\caption{Coverage probability (CP) and average length (AL) of $95\%$ credible intervals of $F_{it}(a_k)$ for $a_k\in \{1,2,4,6,8,10,14\}$, averaged over 100 replicated datasets.
CP is represented by $\%$ scale. 
\label{tab:sim}}
\centering
\medskip
\begin{tabular}{cccccccccccc}
\hline
&& \multicolumn{2}{c}{CP {\footnotesize (Scenario 1)}} & \multicolumn{2}{c}{AL {\footnotesize (Scenario 1)}} & \multicolumn{2}{c}{CP {\footnotesize (Scenario 2)}} & \multicolumn{2}{c}{AL {\footnotesize (Scenario 2)}} \\
Threshold ($a_k$) &  & BIB & BN & BIB & BN & BIB & BN & BIB & BN \\
\hline
1 &  & 91.1 & 7.0 & 0.28 & 0.05 & 93.3 & 7.9 & 0.29 & 0.04 \\
2 &  & 92.8 & 11.1 & 0.30 & 0.06 & 93.0 & 13.0 & 0.32 & 0.06 \\
4 &  & 93.1 & 14.8 & 0.32 & 0.08 & 90.1 & 15.1 & 0.35 & 0.08 \\
5 &  & 93.7 & 16.2 & 0.33 & 0.09 & 89.1 & 15.9 & 0.37 & 0.08 \\
8 &  & 93.8 & 16.6 & 0.33 & 0.09 & 89.7 & 16.6 & 0.38 & 0.09 \\
10 &  & 94.0 & 16.2 & 0.34 & 0.09 & 90.2 & 16.5 & 0.37 & 0.09 \\
14 &  & 93.7 & 15.2 & 0.34 & 0.09 & 90.5 & 15.9 & 0.37 & 0.08 \\
\hline
\end{tabular}
\end{table}

%----------------------------------------%
%          Concluding Remarks            %
%----------------------------------------%
\section{Concluding Remarks} \label{sec:concl} 
In this paper, we have developed a robust spatio-temporal distributional regression (RSTDR) framework by extending boundary-inflated binomial models to spatio-temporal settings to address boundary inflation problems that arise when applying distribution regression in spatio-temporal contexts. 
The proposed dynamic Gaussian predictive process BIB (DGPP-BIB) model combines the flexibility of boundary-inflated binomial modeling with dynamic Gaussian predictive processes that handle irregularly spaced sampling locations varying over time. 
We develop a computationally efficient Bayesian inference algorithm using P\'olya-Gamma data augmentation and dynamic Gaussian predictive processes within a Metropolis-within-Gibbs MCMC framework, enabling scalable posterior computation for the complex hierarchical model structure. 
Extensive simulation studies demonstrate that our approach significantly outperforms existing methods based on standard binomial models, generalized additive models, and gradient boosting approaches. 
The proposed method consistently achieved lower mean squared errors and provided accurate uncertainty quantification with coverage probabilities close to nominal levels, while standard approaches considerably underestimated uncertainty and showed poor coverage properties. 

While our current framework provides substantial improvements over existing methods, promising directions remain for future research. 
Alternative spatio-temporal modeling approaches could be explored. 
Although we used dynamic Gaussian predictive processes for computational feasibility, such low-rank approximations are known to perform poorly in some cases \citep{stein2014limitations}. 
A natural alternative would be to incorporate the Vecchia approximation \citep{vecchia1988estimation} such as nearest neighbor Gaussian process \citep{datta2016hierarchical} into the dynamic model instead of the Gaussian predictive process. 
However, it is difficult to do so when the spatial location varies at each time. 
Therefore, spatio-temporal Gaussian processes with efficient approximation methods \citep{datta2016hierarchical, kang2023correlation} that enable continuous modeling in both space and time dimensions represent a promising future direction, though adapting these approaches to the boundary-inflated setting remains a challenging research problem. 

Additionally, the extension to continuous proportional models could be explored in several directions. 
For instance, \cite{li2018efficient} proposed boundary-inflated beta regression models for non-spatial settings, which could be extended to spatio-temporal contexts to address similar issues for continuous proportional data with support on $[0,1]$. 
\cite{lee2025scalable} developed continuous binomial regression models with spatial extensions, which could be further extended to handle structural boundary inflation in spatio-temporal settings or adapted for sampling locations that vary over time. 
However, an important consideration is the statistical efficiency loss when converting binomial data $y_{it}$ to discrete proportional data $y_{it}/n_{it}$ for continuous proportional models. 
This conversion discards the sample size information $n_{it}$, which is crucial for proper variance modeling and weighting in statistical inference. 
Observations with larger $n_{it}$ provide more reliable proportion estimates, but this reliability information is lost in the conversion process. 
Future research should investigate the statistical trade-offs between direct binomial modeling approaches and conversion-based continuous methods to determine the most appropriate framework for different application settings.

\section*{Acknowledgement}
This work is partially supported by JSPS KAKENHI Grant Numbers 24K21420, 25K21167 and 25H00546.

%=\appendix==========================================
\def\thesection{Appendix}
\def\thesubsection{A.\arabic{subsection}}
\section{}
\def\theequation{A.\arabic{equation}}%%数式の番号付け(section, 式番号)
\def\thethm{A.\arabic{thm}}%%数式の番号付け(section, 式番号)
\def\thelem{A.\arabic{lem}}%%数式の番号付け(section, 式番号)

\setcounter{equation}{0}

\subsection{Block tridiagonal structure from DGPP formulation} \label{app:dgpp} 
One of the key computational advantages of our DGPP-BIB model stems from the transformation of the dynamic Gaussian predictive process formulation into a multivariate normal distribution with block tridiagonal precision matrix structure. 
This transformation enables efficient simulation smoothing algorithms for sampling spatial components $\bar{\bm{u}}$ and ${\bar{\bm{\xi}}_k}$. 

The fundamental insight underlying this transformation is that equation \eqref{eq:dgpp} can be expressed as 
\begin{equation*}
    (\bar{\bu}_1, \bar{\bu}_2, \ldots, \bar{\bu}_T)^\top = (\bar{\bu}_0, \bar{\bu}_1, \ldots, \bar{\bu}_{T-1})^\top + (\bm{e}, \bm{e}, \ldots, \bm{e})^\top, 
\end{equation*} 
where $\bm{e} \sim N_M(\bm{0}_M, \bar{\bC}(\phi_u))$. 
Multiplying both sides by the $T \times T$ matrix $\bH$, which has 1s on the main diagonal and $-1$s on the subdiagonal, yields 
\begin{equation*}
    \bH (\bar{\bu}_1, \bar{\bu}_2, \ldots, \bar{\bu}_T)^\top = (\bar{\bu}_0, \bm{0}_M, \ldots, \bm{0}_M)^\top + (\bm{e}, \bm{e}, \ldots, \bm{e})^\top, 
\end{equation*}
where $\bm{0}_M$ is an $M$-dimensional vector with all elements equal to zero. 
Noting that $\bH$ is the lower triangular matrix and invertible since $\det(\bH) = 1 \neq 0$, we can solve the triangular system $\bH\bm{a} = (\bar{\bm{u}}_0, \bm{0}_M, \ldots, \bm{0}_M)^\top$ for $\bm{a} = (\bar{\bm{a}}_1, \bar{\bm{a}}_2, \ldots, \bar{\bm{a}}_T)^\top$ using forward substitution. 
Specifically, the first equation is $\bar{\bm{a}}_1 = \bar{\bm{u}}_0$, the second equation is $-\bar{\bm{a}}_1 + \bar{\bm{a}}_2 = \bm{0}_M \Rightarrow \bar{\bm{a}}_2 = \bar{\bm{a}}_1 = \bar{\bm{u}}_0$, and more generally, the $t$-th equation is $-\bar{\bm{a}}_{t-1} + \bar{\bm{a}}_t = \bm{0}_M \Rightarrow \bar{\bm{a}}_t = \bar{\bm{a}}_{t-1} = \bar{\bm{u}}_0$. 
Therefore, 
\begin{equation*}
    \bH^{-1}(\bar{\bm{u}}_0, \bm{0}_M, \ldots, \bm{0}_M)^\top = (\bar{\bm{u}}_0, \bar{\bm{u}}_0, \ldots, \bar{\bm{u}}_0)^\top. 
\end{equation*}

Combining this result with the original relationship, we obtain 
\begin{equation*}
    (\bar{\bm{u}}_1, \bar{\bm{u}}_2, \ldots, \bar{\bm{u}}_T) = (\bar{\bm{u}}_0, \bar{\bm{u}}_0, \ldots, \bar{\bm{u}}_0) + (\bm{e}, \bm{e}, \ldots, \bm{e}) (\bH^{-1})^\top
\end{equation*}
Since $\bm{e} \sim N_M(\bm{0}_M, \bar{\bC}(\phi_u))$, the $M \times T$ matrix $(\bm{e}, \bm{e}, \ldots, \bm{e})$ follows a matrix normal distribution $MN_{M \times T}(\bm{0}_{M \times T}, \bar{\bC}(\phi_u), \bm{I}_T)$, where $\bm{0}_{M \times T}$ is the $M \times T$ zero matrix, $\bm{I}_T$ is the $T \times T$ identity matrix, and $\bm{X} \sim MN_{n \times p}(\bm{M}, \bm{U}, \bm{V})$ if and only if its probability density function is 
\begin{equation*}
    (2\pi)^{-np/2} |\bm{V}|^{-n/2} |\bm{U}|^{-p/2} \exp\left\{-\frac{1}{2} \tr(\bm{U}^{-1}(\bm{X}-\bm{M})\bm{V}^{-1}(\bm{X}-\bm{M})^\top)\right\} 
\end{equation*} 
with a $n \times p$ real matrix $\bm{M}$, an $n \times n$ positive-definite matrix $\bm{U}$, and a $p \times p$ positive-definite matrix $\bm{V}$. 
Vectorizing this expression and using the property that $\text{vec}(\bm{X}) \sim N_{np}(\text{vec}(\bm{M}), \bm{V} \otimes \bm{U})$ for $\bm{X} \sim MN_{n \times p}(\bm{M}, \bm{U}, \bm{V})$, we establish the key equivalence 
\begin{equation*}
    \prod_{t=1}^T \phi_M(\bar{\bm{u}}_t; \bar{\bm{u}}_{t-1}, \tau_{u}^{-1} \bar{\bC}(\phi_u)) = \phi_{MT}(\bar{\bm{u}};\bar{\bm{u}}_{0MT},\tau_u^{-1}\bar{\bC}_H)
\end{equation*}
where $\bar{\bC}_H = (\bH^\top \bH)^{-1} \otimes \bar{\bC}(\phi_u)$ and $\otimes$ is the Kronecker product. 
The tridiagonal matrix $\bH^\top \bH$, 
\begin{equation*}
    \bH^\top \bH = \begin{pmatrix}
    2 & -1 & 0 & \cdots & 0 \\
    -1 & 2 & -1 & \ddots & \vdots \\
    0 & -1 & \ddots & \ddots & 0 \\
    \vdots & \ddots & \ddots & 2 & -1 \\
    0 & \cdots & 0 & -1 & 1
    \end{pmatrix}, 
\end{equation*}
and the Kronecker product yield a block tridiagonal precision matrix 
\begin{equation*}
    \tau_u (\bar{\bC}_H)^{-1} = \tau_u (\bH^\top \bH) \otimes \bar{\bC}(\phi_u)^{-1}. 
\end{equation*} 

The same transformation applies to $\{\bar{\bm{\xi}}_k\}$, where the equivalent multivariate normal form is 
\begin{equation*}
    \prod_{t=1}^T \phi_M(\bar{\bm{\xi}}_{kt}; \bar{\bm{\xi}}_{k,t-1}, \tau_{\xi_k}^{-1} \bar{\bC}(\phi_{\xi_k})) = \phi_{MT}(\bar{\bm{\xi}}_k;\bar{\bm{\xi}}_{k0MT},\tau_{\xi_k}^{-1}\bar{\bC}_{Hk})
\end{equation*}
where $\bar{\bC}_{H_k} = (\bH^\top \bH)^{-1} \otimes \bar{\bC}(\phi_{\xi_k})$, yielding analogous block tridiagonal precision matrix structures.

\subsection{Derivative of full conditional posteriors} \label{app:derivative}

We first note that 
\begin{equation*}
    \Bin(y_{it};n_{it},\pi_{it}) = 2^{-n_{it}} \exp(\kappa_{it} \eta_{it}) \int_0^\infty \exp\left( -\frac{1}{2} \omega_{it} \eta_{it}^2 \right) p_{\rm PG}(\omega_{it};n_{it},0) d\omega_{it}, 
\end{equation*}
where $\omega_{it}$ is an additional latent variable for the mixture representation, $\kappa_{it}=y_{it}-n_{it}/2$, $\eta_{it}=\bm{x}_{it}^\top \bbe + u_{it}$, and $p_{\rm PG}(\cdot;b,c)$ is the density function of the P\'olya-gamma distribution \citep{polson2013bayesian}. 
The above integral expression shows that the conditional distribution of $\eta_{it}$ given $\omega_{it}$ is Gaussian, which leads to a tractable posterior computation algorithm. 
The full conditional posterior of $\bbe$ is 
\begin{align*}
    p(\bbe | -)
    &\propto \phi_q(\bbe;\bb_0,\bB_0^{-1}) \exp\left[ -\frac{1}{2} \left\{ \bbe^\top \bX_*^\top \bOm_* \bX_* \bbe - 2\left( \bka_*^\top - \bar{\bu}^\top \bar{\bD}_*^\top \bOm_* \right) \bX_* \bbe \right\} \right] \\
    &\propto \phi_q(\bbe;\bB^{-1}\bb,\bB^{-1}),
\end{align*}
where $\bb=\bB_0\bb_0+\bX_*^\top(\bka_*-\bOm_*\bu_*)$ and $\bB=\bB_0+\bX_*^\top\bOm_*\bX_*$. 
The full conditional posterior of $\bar{\bu}$ is 
\begin{equation} \label{eq:u}
    \begin{split}
        p(\bar{\bu}|-)
        \propto& \prod_{t=1}^T \phi_M(\bar{\bm{u}}_t;\bar{\bm{u}}_{t-1},\tau_{u}^{-1} \bar{\bC}(\phi_u)) \\
        &\times \exp\left[ -\frac{1}{2} \left\{ \bar{\bu}^\top\bar{\bD}_*^\top\bOm_*\bar{\bD}_*\bar{\bu} - 2\left( \bka_*^\top - \bbe^\top\bX_*^\top\bOm_* \right) \bar{\bD}_*^\top \bar{\bu} \right\} \right] \\
        =& \phi_{MT}(\bar{\bu};\bar{\bu}_{0MT},\tau_u^{-1}\bar{\bC}_H) \\
        &\times \exp\left[ -\frac{1}{2} \left\{ \bar{\bu}^\top\bar{\bD}_*^\top\bOm_*\bar{\bD}_*\bar{\bu} - 2\left( \bka_*^\top - \bbe^\top\bX_*^\top\bOm_* \right) \bar{\bD}_*^\top \bar{\bu} \right\} \right] \\
        \propto& \phi_{MT}(\bar{\bu};\bQ^{-1}\bem,\bQ^{-1}),
    \end{split}
\end{equation}
where $\bem=\tau_u\bar{\bC}_H^{-1}\bar{\bu}_{0MT}+\bar{\bD}_*^\top(\bka_*-\bOm_*\bX_*\bbe)$, $\bQ=\tau_u\bar{\bC}_H^{-1}+\bar{\bD}_*^\top\bOm_*\bar{\bD}_*$, $\bar{\bu}_{0MT}$ is an $MT \times 1$ vector constructed by vertically stacking $T$ copies of $\bar{\bu}_0$, and $\bar{\bC}_H=(\bH^\top\bH)^{-1}\otimes\bar{\bC}(\phi_u)$ with the $T\times T$ matrix $\bH$ having 1s on the main diagonal and $-1$s on the subdiagonal. 
Equivalently, $\bem=(\bem_1^\top,\ldots,\bem_T^\top)^\top$ with $\bem_1=\tau_u \bar{\bC}(\phi_u)^{-1} \bar{\bu}_0+\bar{\bD}_{1*}^\top (\bka_{1*}-\bOm_{1*}\bX_{1*}\bbe)$ and $\bem_t=\bar{\bD}_{t*}^\top (\bka_{t*}-\bOm_{t*}\bX_{t*}\bbe)$ ($t=2,\ldots,T$), and
\begin{equation*}
    \bQ = 
    \begin{pmatrix}
        \bQ_{11} & \bQ_{12} & \bm{0}   & \cdots        & \bm{0} \\
        \bQ_{21} & \bQ_{22} & \bQ_{23} & \ddots        & \vdots \\
        \bm{0}   & \bQ_{32} & \ddots   & \ddots        & \bm{0} \\
        \vdots   & \ddots   & \ddots   & \bQ_{T-1,T-1} & \bQ_{T-1,T} \\
        \bm{0}   & \cdots   & \bm{0}   & \bQ_{T,T-1}   & \bQ_{TT}
    \end{pmatrix}
\end{equation*}
with $\bQ_{tt} = 2\tau_u\bar{\bC}(\phi_u)^{-1}+\bar{\bD}_{t*}^\top \bOm_{t*} \bar{\bD}_{t*}$, $\bQ_{t,t+1}=-\tau_u\bar{\bC}(\phi_u)^{-1}$ ($t=1,\ldots,T-1$) and $\bQ_{TT} = \tau_u\bar{\bC}(\phi_u)^{-1}+\bar{\bD}_{T*}^\top \bOm_{T*} \bar{\bD}_{T*}$. 
This block tridiagonal structure enables efficient simulation smoothing algorithms (see Appendix \ref{app:spatial}). 
Note that the conversion from the first to the second line in Equation \eqref{eq:u} employs the DGPP transformation detailed in Appendix \ref{app:dgpp}. 
The full conditional posterior of $\{\omega_{it}\}$ is the P\'olya-Gamma distribution $p_{\PG}(\omega_{it};n_{it},\eta_{it})$ from Theorem 1 in \cite{polson2013bayesian}. 

Moreover, since 
\begin{equation*}
    \frac{e^{\psi_{kit}}}{1+e^{\psi_{0it}}+e^{\psi_{1it}}} = \frac{\exp(\psi_{kit}-\Psi_{kit})}{1+\exp(\psi_{kit}-\Psi_{kit})}, 
\end{equation*}
where $\Psi_{kit} = \log\{1+\sum_{\ell\in \{0,1\}\setminus \{k\}}\exp(\psi_{\ell it})\}$, 
\begin{align*}
    \frac{e^{{\psi_{kit}}^{I(r_{it}=k)}}}{1+e^{\psi_{0it}}+e^{\psi_{1it}}} 
    \propto& \exp\{\kappa_{kit}(\psi_{kit}-\Psi_{kit})\} \\
    &\times \int_0^\infty \exp\left\{ -\frac{1}{2} \omega_{kit} (\psi_{kit}-\Psi_{kit})^2 \right\} p_{\PG}(\omega_{kit};1,0) d\omega_{kit},
\end{align*}
where $\kappa_{kit}=I(r_{it}=k)-1/2$. 
Therefore, the above integral expression also shows that the conditional distribution of $\psi_{kit}$ given $\omega_{kit}$ is Gaussian. 
The full conditional posterior of $\bga_k$ is 
\begin{align*}
    p(\bga_k|-) 
    \propto& \phi_q(\bga_k;\bg_{k0},\bG_{k0}^{-1}) \\
    &\times \exp\left[ -\frac{1}{2} \left\{ \bga_k^\top \bX^\top \bOm_k \bX \bga_k - 2\left( \bka_k^\top + \bPs_k^\top \bOm_k - \bar{\bxi}_k^\top \bar{\bD}_k^\top \bOm_k \right) \bX \bga_k \right\} \right] \\
    \propto& \phi_q(\bga_k; \bG_k^{-1} \bg_k, \bG_k^{-1}),
\end{align*}
where $\bg_k = \bG_{k0}\bg_{k0} + \bX^\top \{\bka_k + \bOm_k(\bPs_k - \bxi_k)\}$ and $\bG_k = \bG_{k0} + \bX^\top \bOm_k \bX$. 
The full conditional posterior of $\bar{\bxi}_k$ is 
\begin{align*}
    p(\bar{\bxi}_k|-) 
    \propto& \phi_{MT}(\bar{\bxi}_k; \bar{\bxi}_{k0MT}, \tau_{\xi_k}^{-1} \bar{\bC}_{H_k}) \\
    &\times \exp\left[ -\frac{1}{2} \left\{ \bar{\bxi}_k^\top \bar{\bD}_k^\top \bOm_k \bar{\bD}_k \bar{\bxi}_k -2\left( \bka_k^\top + \bPs_k^\top \bOm_k - \bga_k^\top \bX^\top \bOm_k \right) \bar{\bD}_k \bar{\bxi}_k \right\} \right] \\
    \propto& \phi_{MT}(\bar{\bxi}_k; \bQ_k^{-1} \bem_k, \bQ_k^{-1}),
\end{align*}
where $\bem_k = \tau_{\xi_k} \bar{\bC}_{H_k}^{-1} \bar{\bxi}_{k0MT} + \bar{\bD}_k^\top \{\bka_k + \bOm_k (\bPs_k - \bX \bga_k)\}$, $\bQ_k = \tau_{\xi_k} \bar{\bC}_{H_k}^{-1} + \bar{\bD}_k^\top \bOm_k \bar{\bD}_k$, $\bar{\bxi}_{k0MT}$ is an $MT \times 1$ vector constructed by vertically stacking $T$ copies of $\bar{\bxi}_{k0}$, and $\bar{\bC}_{H_k}=(\bH^\top\bH)^{-1}\otimes\bar{\bC}(\phi_{\xi_k})$. 
Equivalently, $\bem_k = (\bem_{k1}^\top,\ldots,\bem_{kT}^\top)^\top$ with $\bem_{k1} = \tau_{\xi_k} \bar{\bC}(\phi_{\xi_k})^{-1} \bar{\bxi}_{k0} + \bar{\bD}_{k1}^\top (\bar{\bka}_{k1}+\bOm_{k1}\bPs_{k1}-\bOm_{k1}\bX_1\bga_k)$ and $\bem_{kt} = \bar{\bD}_{kt}^\top (\bar{\bka}_{kt}+\bOm_{kt}\bPs_{kt}-\bOm_{kt}\bX_t\bga_k)$ ($t=2,\ldots,T$), and 
\begin{equation*}
    \bQ_k = 
    \begin{pmatrix}
        \bQ_{k11} & \bQ_{k12} & \bm{0}     & \cdots          & \bm{0} \\
        \bQ_{k21} & \bQ_{k22} & \bQ_{k23}  & \ddots          & \vdots \\
        \bm{0}    & \bQ_{k32} & \ddots     & \ddots          & \bm{0} \\
        \vdots    & \ddots    & \ddots     & \bQ_{k,T-1,T-1} & \bQ_{k,T-1,T} \\
        \bm{0}     & \cdots     & \bm{0}   & \bQ_{k,T,T-1}   & \bQ_{kTT}
    \end{pmatrix}
\end{equation*}
with $\bQ_{ktt} = 2\tau_{\xi_k}\bar{\bC}(\phi_{\xi_k})^{-1}+\bar{\bD}_{kt}^\top \bOm_{kt} \bar{\bD}_{kt}$, $\bQ_{k,t,t+1}=-\tau_{\xi_k}\bar{\bC}(\phi_{\xi_k})^{-1}$ ($t=1,\ldots,T-1$) and $\bQ_{kTT} = \tau_{\xi_k}\bar{\bC}(\phi_{\xi_k})^{-1}+\bar{\bD}_{kT}^\top \bOm_{kT} \bar{\bD}_{kT}$. 
Similarly to sampling $\bar{\bu}$, the simulation smoothing described previously enables efficient sampling of $\{\bar{\bxi}_k\}$. 
As with $\bar{\bu}$, this block tridiagonal structure enables efficient simulation smoothing algorithms for sampling $\bar{\bxi}_k$. 
The full conditional posterior of $\{\omega_{kit}\}$ is also the P\'olya-Gamma distribution $p_{\PG}(\omega_{kit};1,\psi_{kit}-\Psi_{kit})$. 

The full conditional posterior of the latent indicator variable $\{r_{it}\}$ in \eqref{eq:latent} is the categorical distribution, that is, 
\begin{equation*}
    \Pr(r_{it}=k|-) = 
    \begin{cases}
        \tilde{p}_{0it} = \dfrac{e^{\psi_{0it}}\delta_0(y_{it})}{e^{\psi_{0it}}\delta_0(y_{it}) + e^{\psi_{1it}}\delta_{n_{it}}(y_{it}) + \Bin(y_{it};n_{it},\pi_{it})} & (k=0) \\
        \tilde{p}_{1it} = \dfrac{e^{\psi_{1it}}\delta_{n_{it}}(y_{it})}{e^{\psi_{0it}}\delta_0(y_{it}) + e^{\psi_{1it}}\delta_{n_{it}}(y_{it}) + \Bin(y_{it};n_{it},\pi_{it})} & (k=1) \\
        1-\tilde{p}_{0it}-\tilde{p}_{1it} & (k=2)
    \end{cases},
\end{equation*}
and the full conditional posteriors of the initial states of the spatial components $\{\bar{\bu}_0, \bar{\bxi}_{k0}\}$ and the remaining spatial parameters $\{\tau_u, \tau_{\xi_k}, \phi_u, \phi_{\xi_k}\}$ are as follows: 
\begin{align*}
    p(\bar{\bu}_0 |-) 
    &\propto \phi_M(\bar{\bu}_0; \bm{0}, \tau_{u}^{-1} \bar{\bC}(\phi_u)) \phi_M(\bar{\bu}_1; \bar{\bu}_0, \tau_{u}^{-1} \bar{\bC}(\phi_u)) \\
    &\propto \phi_M(\bar{\bu}_0; 2^{-1}\bar{\bu}_1, (2\tau_{u})^{-1} \bar{\bC}(\phi_u)), \\
    p(\bar{\bxi}_{k0} |-) 
    &\propto \phi_M(\bar{\bxi}_{k0}; \bm{0}, \tau_{\xi_k}^{-1} \bar{\bC}(\phi_{\xi_k})) \phi_M(\bar{\bxi}_{k1}; \bar{\bxi}_{k0}, \tau_{\xi_k}^{-1} \bar{\bC}(\phi_{\xi_k})) \\
    &\propto \phi_M(\bar{\bxi}_{k0}; 2^{-1}\bar{\bxi}_{k1}, (2\tau_{\xi_k})^{-1} \bar{\bC}(\phi_{\xi_k})), 
\end{align*}
\begin{align*}
    p(\tau_{u} |-) 
    \propto& \tau_{u}^{a_{u_0}-1} e^{-b_{u_0} \tau_{u}} \phi_M(\bar{\bu}_0; \bm{0}, \tau_{u}^{-1} \bar{\bC}(\phi_u)) \phi_{MT}(\bar{\bu};\bar{\bu}_{0MT},\tau_u^{-1}\bar{\bC}_H) \\
    \propto& \tau_{u}^{a_{u_0} + M(T+1)/2 -1} \\ 
    &\times \exp \left[ - \tau_{u} \left\{ b_{u_0} + \frac12 \bar{\bu}_0^\top \bar{\bC}(\phi_u)^{-1} \bar{\bu}_0 + \frac12 (\bar{\bu}-\bar{\bu}_{0MT})^\top \bar{\bC}_H^{-1} (\bar{\bu}-\bar{\bu}_{0MT}) \right\}  \right] \\
    \propto& {\rm Ga}(\tau_{u}; a_u, b_u), 
\end{align*}
where
\begin{align*}
    a_u &= a_{u_0} + \frac{M(T+1)}{2}, \\ 
    b_u &= b_{u_0} + \frac{1}{2}\left[ \bar{\bu}_0^\top \bar{\bC}(\phi_u)^{-1} \bar{\bu}_0 + \tr( \bm{U}^\top \bar{\bC}(\phi_u)^{-1} \bm{U} \bH^\top \bH ) \right], 
\end{align*}
${\rm vec}(\bm{U}) = \bar{\bu}-\bar{\bu}_{0MT}$, and $\tr(\bm{A})$ is the trace of a square matrix $\bm{A}$, 
\begin{align*}
    p(\tau_{\xi_k} |-) 
    \propto& \tau_{\xi_k}^{a_{\xi_{k0}}-1} e^{-b_{\xi_{k0}} \tau_{\xi_k}} \phi_M(\bar{\bxi}_{k0}; \bm{0}, \tau_{\xi_{k}}^{-1} \bar{\bC}(\phi_{\xi_{k}})) \phi_{MT}(\bar{\bxi}_k;\bar{\bxi}_{k0MT},\tau_{\xi_{k}}^{-1}\bar{\bC}_{H_k}) \\
    \propto& \tau_{\xi_{k}}^{a_{\xi_{k0}} + M(T+1)/2 -1} \\ 
    &\times \exp \left[ - \tau_{\xi_{k}} \left\{ b_{\xi_{k0}} + \frac12 \bar{\bxi}_{k0}^\top \bar{\bC}(\phi_{\xi_{k}})^{-1} \bar{\bxi}_{k0} + \frac12 (\bar{\bxi}_k-\bar{\bxi}_{k0MT})^\top \bar{\bC}_{H_k}^{-1} (\bar{\bxi}_k-\bar{\bxi}_{k0MT}) \right\}  \right] \\
    \propto& {\rm Ga}(\tau_{\xi_{k}}; a_{\xi_{k}}, b_{\xi_{k}}), 
\end{align*}
where 
\begin{align*}
    a_{\xi_k} &= a_{\xi_{k0}} + \frac{M(T+1)}{2}, \\
    b_{\xi_k} &= b_{\xi_{k0}} + \frac{1}{2}\left[ \bar{\bxi}_{k0}^\top \bar{\bC}(\phi_{\xi_k})^{-1} \bar{\bxi}_{k0} + \tr( \bm{\Xi}^\top \bar{\bC}(\phi_{\xi_k})^{-1} \bm{\Xi} \bH^\top \bH ) \right],
\end{align*}
and ${\rm vec}(\bm{\Xi}) = \bar{\bxi}_k-\bar{\bxi}_{k0MT}$, and 
\begin{align*}
    p(\phi_u|-) 
    \propto& \prod_{t=1}^T \prod_{i=1}^{N_t} \Bin(y_{it};n_{it},\pi_{it})^{I(r_{it}=2)} \\
    &\times \phi_M(\bar{\bu}_0; \bm{0}, \tau_{u}^{-1} \bar{\bC}(\phi_u)) \prod_{t=1}^T \phi_M(\bar{\bm{u}}_t; \bar{\bm{u}}_{t-1}, \tau_{u}^{-1} \bar{\bC}(\phi_u)), \\
    p(\phi_{\xi_k}|-) 
    \propto& \prod_{t=1}^T\prod_{i=1}^{N_t}
p_{kit}^{I(r_{it}=k)} \\
    &\times \phi_M(\bar{\bm{\xi}}_{k0}; \bm{0}, \tau_{\xi_k}^{-1} \bar{\bC}(\phi_{\xi_k})) \prod_{t=1}^T \phi_M(\bar{\bm{\xi}}_{kt}; \bar{\bm{\xi}}_{kt-1}, \tau_{\xi_k}^{-1} \bar{\bC}(\phi_{\xi_k})).
\end{align*}

\subsection{Sampling spatial components} \label{app:spatial}
The block tridiagonal precision matrix structures derived in Appendix A.2 enable efficient simulation smoothing algorithms for sampling both $\bar{\bu}$ and $\bar{\bxi}_k$. 
We employ two computationally efficient approaches that exploit these structures and avoid the computational bottlenecks of traditional Kalman filter-based approaches. 

The first approach is the Cholesky factor algorithm of \cite{rue2001fast}, which directly factorizes the precision matrix $\bQ$ using its band diagonal structure. 
The algorithm proceeds as follows. 
\begin{enumerate}
    \item 
    Compute the band Cholesky decomposition $\bQ = \bm{L}\bm{L}^\top$. 

    \item 
    Solve $\bm{L}\bm{v} = \bem$ using forward substitution. 

    \item 
    Generate $\bm{\ep} \sim N_{MT}(\bm{0}, \bm{I}_{MT})$ and solve $\bm{L}^\top\bar{\bu} = \bm{v} + \bm{\ep}$ using backward substitution. 
\end{enumerate}

The second approach is the simulation smoothing algorithm developed by \cite{mccausland2011simulation}, which is specifically designed for block tridiagonal structures in normal linear state-space models. 
The sampling procedure is as follows and consists of two steps. 
\begin{enumerate}
    \item 
    For $t=1,\ldots,T$, 
    \begin{enumerate}
        \item 
        If $t=1$, then set $\bSi_1^{-1} = \bQ_{11}$; otherwise, set $\bSi_t^{-1} = \bQ_{tt} - (\bQ_{t-1,t}^\top \bSi_{t-1} \bQ_{t-1,t})$. 
        
        \item 
        Compute the Cholesky decomposition $\bSi_t^{-1} = \bLa_t \bLa_t^\top$.

        \item 
        Compute $\bLa_t^{-1} \bQ_{t,t+1}$ using triangular back-substitution. 

        \item 
        Compute $\bQ_{t,t+1}^\top \bSi_t \bQ_{t,t+1} = (\bLa_t^{-1} \bQ_{t,t+1})^\top (\bLa_t^{-1} \bQ_{t,t+1})$. 

        \item 
        If $t=1$, then compute $\bmu_1 = (\bLa_1^\top)^{-1} (\bLa_1^{-1}\bem_1)$; \\
        otherwise, compute $\bmu_t = (\bLa_t^\top)^{-1} [\bLa_t^{-1}(\bem_t-\bQ_{t-1,t}^\top \bmu_{t-1})]$. 
    \end{enumerate}

    \item 
    For $t=T,\ldots,1$, 
    \begin{enumerate}
        \item 
        Sample $\bep_t$ from $N_M(\bm{0},\bm{I}_M)$. 

        \item 
        If $t=T$, then compute $\bar{\bu}_T = \bmu_T + (\bLa_T^\top)^{-1} \bep_T$; \\
        otherwise, compute $\bar{\bu}_t = \bmu_t + (\bLa_t^\top)^{-1} [\bep_t-(\bLa_t^{-1}\bQ_{t,t+1})\bar{\bu}_{t+1}]$. 
    \end{enumerate}
\end{enumerate}

\medskip
\medskip
The same algorithms apply to sampling $\bar{\bxi}_k$ using the corresponding block tridiagonal precision matrices $\bQ_k$ and mean vectors $\bem_k$ defined in Appendix \ref{app:derivative}.

\vspace{1cm}

%   Reference
\bibliographystyle{chicago}
\bibliography{Ref}

\end{document}